**Stable, mode-matched, medium-finesse optical cavity incorporating a microcantilever mirror**

J.G.E. Harris, B. M. Zwickl, A. M. Jayich

*Department of Physics, Yale University, P.O. Box 208284, New Haven CT, 06520*

**Abstract**

A stable optical resonator has been built using a 30 µm-wide, metal-coated microcantilever as one mirror. The second mirror was a 12.7 mm-diameter concave dielectric mirror. By positioning the two mirrors 75 mm apart in a near-hemispherical configuration, a Fabry-Perot cavity with a finesse equal to 55 was achieved. The finesse was limited by the optical loss in the cantilever's metal coating; diffraction losses from the small mirror were negligible. The cavity achieved passive laser cooling of the cantilever's Brownian motion.

The coupling of micromechanical structures to light via radiation pressure has been a topic of considerable interest in recent years. Much of this interest has been driven by the goal of observing and exploiting quantum aspects of either the radiation pressure or the micromechanical structures themselves. Applications to quantum-limited detectors,[1,2] the generation of squeezed[3] and entangled[4] light, and macroscopic quantum phenomena[5] have been proposed. Many of these goals require a high-finesse optical cavity in which one mirror is mounted on a micromechanical structure such as a cantilever. Realizing such a cavity is challenging because the cantilever mirror should be quite small (~10's of μm) and this small size may lead to diffraction losses which will degrade the optical cavity's finesse.

Here we describe the construction and characterization of a Fabry-Perot cavity in which one mirror is formed by a commercial atomic force microscope cantilever. The second mirror is concave and forms a stable optical resonator with the cantilever. This is in contrast to previous experiments which have either used two planar mirrors (leading to an unstable cavity and low finesse)[6,7] or substantially larger (mm-scale) cantilevers.[1,8]

The two mirrors are arranged in a near-hemispherical geometry in order to support cavity modes with the smallest possible waist at the cantilever. We mode-match a stable cw laser to this cavity and predominantly excite a single transverse and longitudinal mode, which we measure to have a finesse $\mathcal{F} = 55$. This value is consistent with the independently measured reflectivites of the two mirrors forming the cavity, indicating that diffraction loss from the small cantilever mirror does not limit $\mathcal{F}$. In addition, we find that the nonadiabatic response of the cantilever to the light in the cavity leads to passive laser cooling of the cantilever's Brownian motion by roughly a factor of six.

The cavity is shown schematically in Fig. 1. It is formed by a microcantilever and a concave dielectric mirror. The cantilever is a commercial AFM cantilever (Olympus AC240TM), 240 μm long, 30 μm wide, and 2.7 μm thick. The cantilever is made of Si. The surface of the cantilever facing the input coupler is coated with Al, while the opposite surface is coated with Pt. It has a measured resonance frequency in its lowest flexural mode of $\nu_0 = 67,110$ Hz, and is specified to have a spring constant $k = 2$ N/m. We measured the reflectivity of the cantilever at $\lambda = 532$ nm to be $R_2 = 0.83 \pm 0.05$.

The dielectric mirror (which serves as the cavity input coupler) is a 12.7 mm-diameter concave mirror (CVI PR1-532-99-0537-0.075cc). Its radius of curvature is specified to be $r = 75$ mm. We measured its reflectivity at $\lambda = 532$ nm to be $R_1 = 0.995$. Thus the cavity is undercoupled.

Throughout this paper we use a coordinate system $\{x, y, z\}$ whose origin $\{0,0,0\}$ is defined to be the center of curvature of the input coupler. The directions of the $x$, $y$, and $z$ axes are shown in Fig. 1.

The dielectric mirror is mounted in a tilt stage (ThorLabs KM05) and affixed to one end of an Invar spacer. The cantilever is mounted on a second tilt stage which in turn is mounted on a piezoelectric stick-slip $x$-$y$-$z$ translation stage (AttoCube ANP100 series). This translation stage provides several mm of travel and sub-micron positioning accuracy in all three axes. The translation stage is mounted to the end of the Invar spacer opposite the dielectric mirror. The spacer is mounted inside a 150 mm-diameter vacuum chamber which provides optical access to both ends of the cavity via antireflection-coated windows.

The cavity is illuminated by $\lambda = 532$ nm light from a doubled cw Nd:YAG laser (Innolight Prometheus). The collimated laser beam is expanded to a $1/e^2$ diameter of $\omega_{in} = 13.5$ mm, and mode matched to the cavity by a 50 mm diameter, 175 mm focal length lens positioned in front of the vacuum chamber.

Since the cantilever mirror is approximately planar all the cavity modes have waists at the cantilever. In order to minimize the diameter of the cavity mode waists, we position the cantilever very close to $\{0,0,0\}$ (i.e., near-hemispherical geometry). Thus the cavity length $L = r = 75$ mm. This is achieved by first aligning the mode-matching lens so that the beam waist is roughly at $\{0,0,0\}$. Using the piezo translation stage the cantilever is adjusted in the $x$-$y$ plane until it partially obscures the beam (as monitored via light transmitted through the cavity). Then the $z$ position of the cantilever is adjusted until it approaches $\{0,0,0\}$. We determine the optimal alignment by scanning the cantilever position through a few free cavity spectral ranges (using the $z$-piezo of the AttoCube) and monitoring the reflected signal.

Figures 2(a)-2(c) show the reflected signal $P_r$ as a function of the displacement $\delta z$ when the cantilever is in different positions along the $z$ axis. At positive displacements along $z$ (i.e., toward the input coupler from $\{0,0,0\}$), the cavity satisfies $0 < (1 - L/r) < 1$ and so is nominally stable[9] (although the diffraction losses may still be large). This is illustrated by the data in Fig. 2(a). As the cantilever is moved closer to {0,0,0} the finesse increases and the transverse mode spacing decreases. The increasing finesse presumably results from the fact that the diffraction losses decrease as the spot size at the cantilever decreases. The decrease of the transverse mode spacing is expected for a cavity as the mirror positions approach a hemispherical geometry, and when the hemispherical geometry is reached, the transverse modes should become degenerate.[10] Figure 2(b) shows a trace of the cavity reflection in which the transverse modes are approximately degenerate. We interpret this as a sign that the cantilever is very close to {0,0,0}, and hence that the cavity is roughly hemispherical.

When the cantilever is translated beyond {0,0,0} (i.e., to negative $z$), the finesse drops rapidly (Fig. 2(c)). This is because $0 < (1 - L/r) < 1$ is no longer satisfied and the cavity is not stable.

When the cantilever is positioned at the optimum $z$ but displaced in the $x$-$y$ plane the finesse also drops. Figure 2(d) shows $P_r(z)$ for the cantilever displaced by approximately 35 μm along its narrowest dimension (i.e., to {0, 35 μm, 0}). Although this displacement implies that a large fraction of the input laser beam does not intersect the cantilever, the finesse still remains relatively high (~10), due to the fact that the cavity modes readjust themselves to minimize diffraction loses.

Figure 3 shows a finer scan of the cantilever position around {0,0,0}. Fitting the data to the expression[9]

$$P_r(z) = 1 - \varepsilon \left(1 + \left(\frac{2\mathcal{F}}{\pi} \sin\left(\frac{2\pi(L + \delta z)}{\lambda}\right)\right)^2\right)^{-1}$$

gives a value for the finesse $\mathcal{F} = 55$. This is roughly consistent with $\pi/(1 - (R_1 R_2)^{1/2})$, the value expected in the absence of diffraction losses.

Although many of the long-term goals for these devices involve quantum effects, micromechanical optical cavities like the ones described here can also exhibit interesting classical effects. One example is optical control of the cantilever's Brownian motion. Optical control of the Brownian motion is achieved when the optical force on the cantilever has a phase lag relative to the cantilever motion. A positive phase lag decreases the cantilever's damping while a negative phase lag increases it. As discussed extensively elsewhere,[11,7] this change in the damping corresponds to a change in the effective temperature of the cantilever.

Figure 4 shows the power spectral density of the cantilever's undriven, Brownian motion. This data is acquired using 633 nm light from a HeNe laser. Since the dielectric mirror's reflectivity at 633 nm is only 0.27, the 633 nm light sees a low-finesse (and overcoupled) cavity. The lower finesse for this laser helps to ensure that it does not perturb the cantilever motion. In addition, only 0.18 mW of 633 nm light is incident on the cavity.

The data indicated by square points were taken with the 532 nm laser shuttered. Fitting this data to the expected form for a damped harmonic oscillator gives a value of the quality factor $Q = 1925$.

The data indicated by circular points were taken with 4.53 mW of 532 nm light incident on the cavity and slightly red-detuned from the cavity resonance. In this data the cantilever $Q$ drops to 370. The area under these curves is proportional to $\langle x^2 \rangle$, the mean-squared amplitude of the cantilever's Brownian motion. With the 532 nm laser on, $\langle x^2 \rangle$ is decreased by a factor of ~6.15. This implies a reduction of the cantilever's effective temperature by the same amount, i.e., from 300 K to ~50 K.

The phase shift responsible for this cooling can have a variety of origins. The two most relevant here are the cavity ringdown time $\tau_{opt} = 2L\mathcal{F}/\pi c$ (which sets the phase lag between the radiation pressure and the cantilever motion) and the cantilever's thermalization time constant $\tau_T$ (which depends upon the thermal conductivity and heat capacity of the cantilever and sets the phase lag due to photothermal forces on the cantilever). From our measurement of $\mathcal{F}$ we find $\tau_{opt} = 7.5$ ns, roughly $2 \times 10^4$ times longer than in Ref. 7. However, simple estimates of the radiation pressure-induced cooling show that this value of $\tau_{opt}$ should lead to cooling of only ~15%. The fact that we observe a factor of six decrease in the cantilever temperature suggests that the majority of this cooling arises from photothermal effects, as in Ref. 7.

We gratefully acknowledge helpful discussions with Steven Girvin, Richard Mirin, and Jeff Thompson.

**Figure Captions**

**Fig. 1:** Scale drawing of the cavity. The cantilever chip is visible at the right-hand side of the drawing; the cantilever itself is not visible at this scale. Also shown is the orientation of the coordinate system used in the text. The coordinate system's origin is at the center of curvature of the input coupler. The force of gravity points along the negative $y$ direction.

**Fig. 2:** Reflected intensity as a function of cantilever displacement for four different cavity geometries. In **a)** the cantilever is positioned at approximately {0, 0, 400 µm}, i.e., 400 µm towards the input coupler from the input coupler's center of curvature (the coordinate system is given Fig. 1). The cavity is nominally stable, but diffraction losses appear to limit the finesse. Multiple transverse modes are visible within each free spectral range. In **b)** the cantilever is approximately at {0, 0, 0}, i.e., the hemispherical configuration. Here the finesse is at its maximum and the transverse modes visible in **a)** have become degenerate. In **c)** the cantilever is at {0, 0, −500 µm}. This renders the cavity unstable and leads to a low finesse. In **d)** the cantilever is approximately at {0, 35 µm, 0}. Here the input beam only partially overlaps with the cantilever.

**Fig. 3:** A more detailed scan of the cavity reflection peak. Here the cantilever is positioned at approximately {0,0,0}, i.e., in the hemispherical geometry. The solid line is a fit to the formula given in the text, and corresponds to a finesse $\mathcal{F} = 55$.

**Fig. 4**: Laser cooling of the cantilever. $S_x$, the power spectral density of the cantilever displacement, is plotted with the 532 nm cooling laser off (squares) and on (circles). The solid lines are fits to the expected form for a damped harmonic oscillator. With the cooling laser off, the fit gives $Q = 1925$. With the cooling laser on, $Q = 370$.

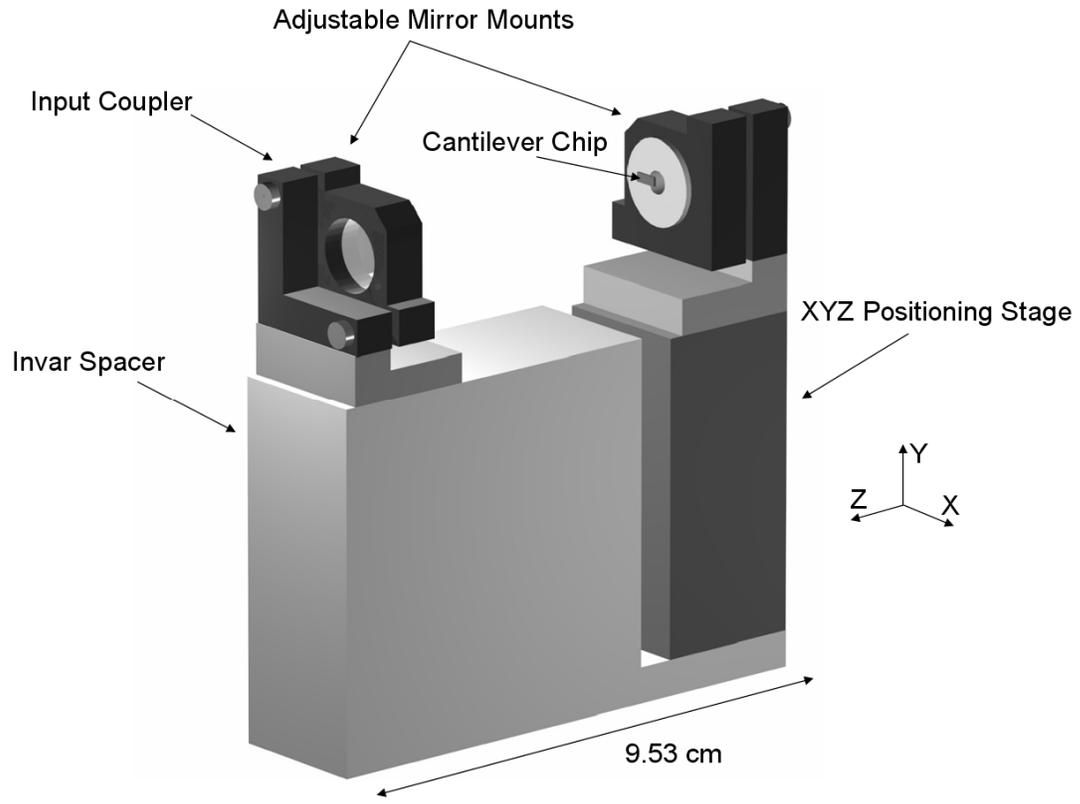

**Figure 1**

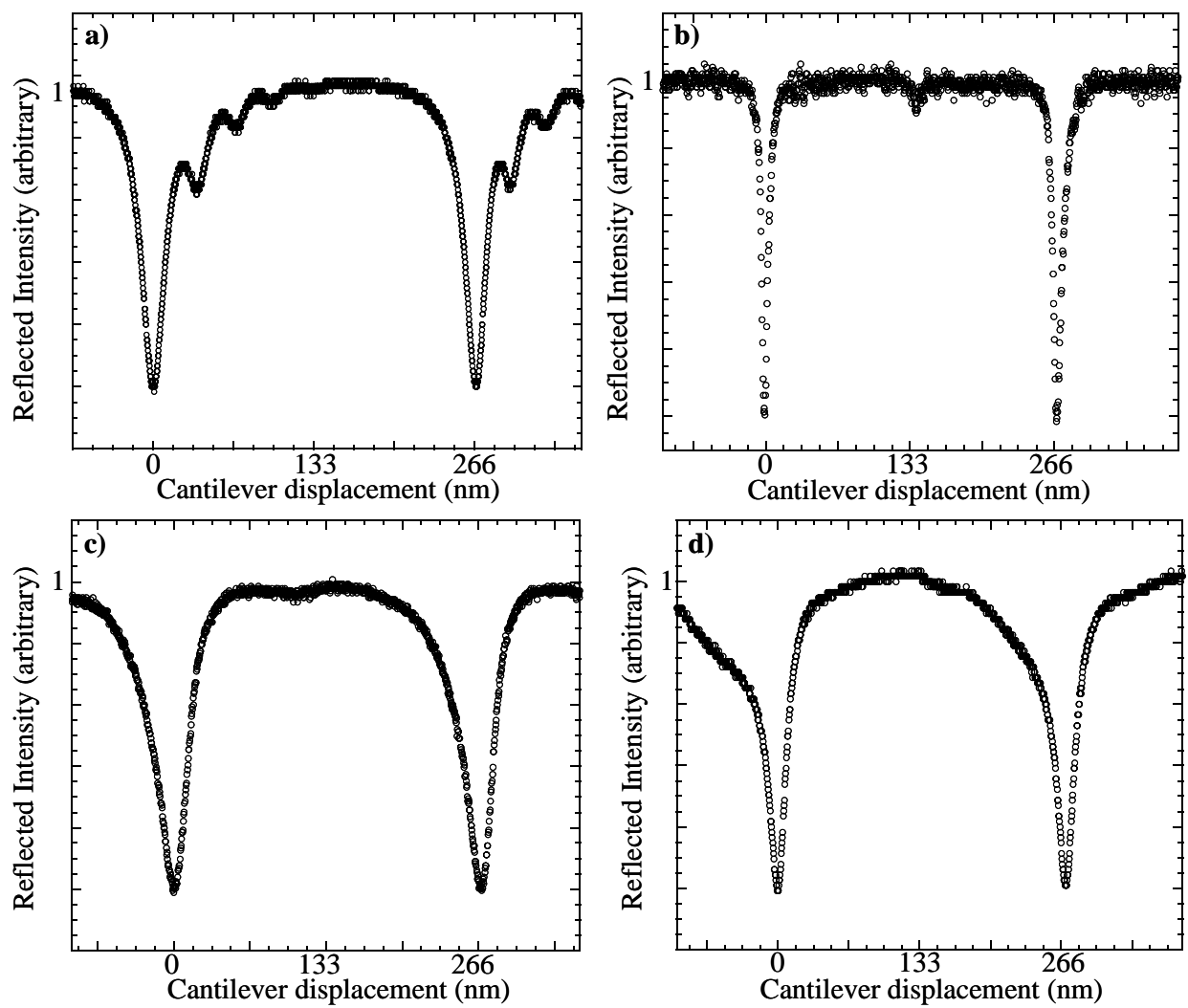

**Figure 2**

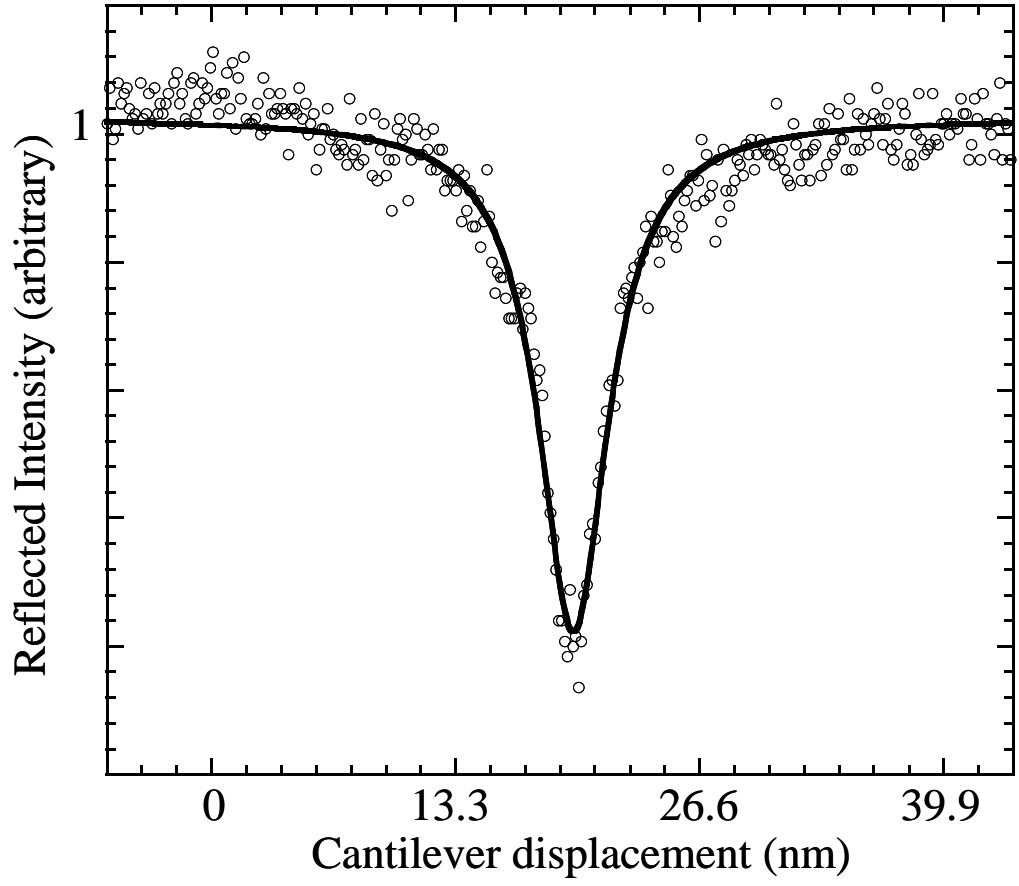

**Figure 3**

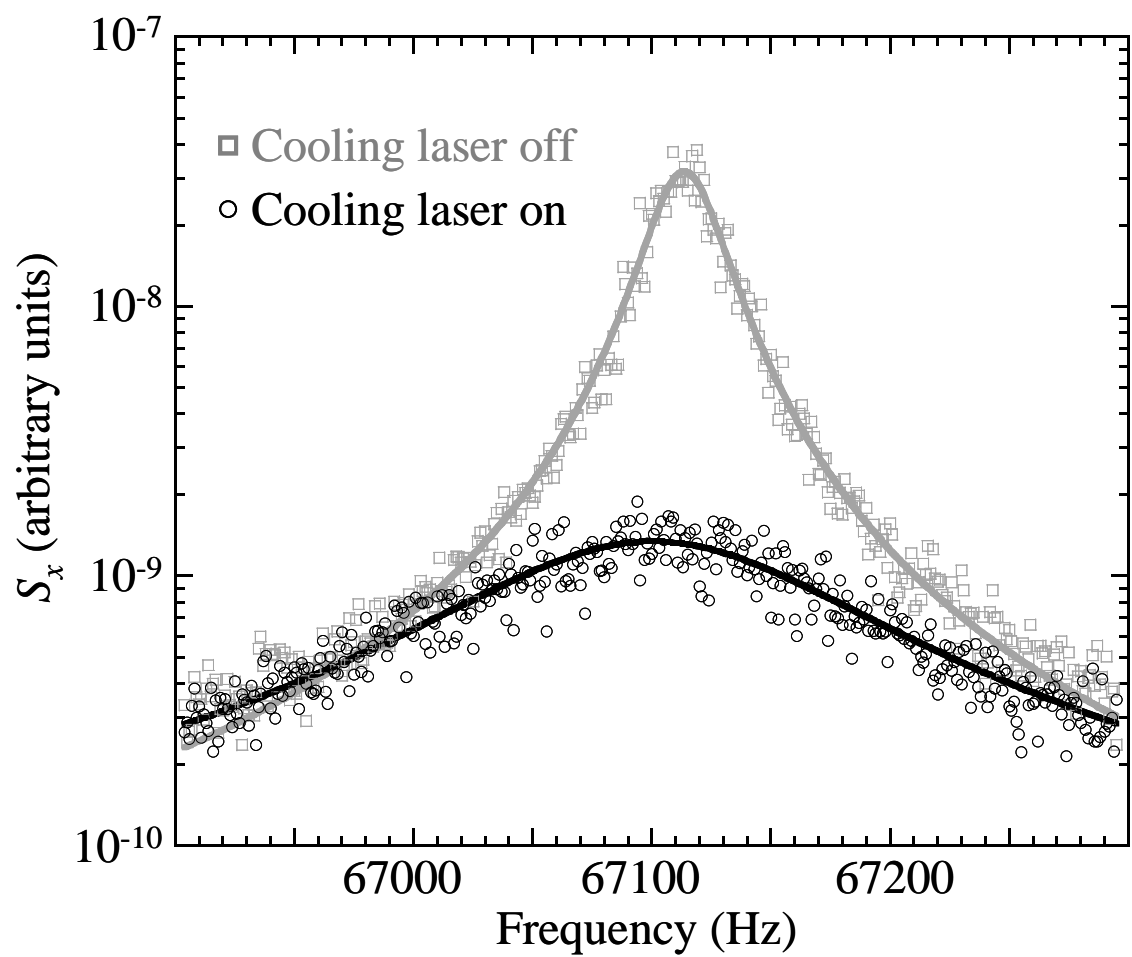

Figure 4